\documentclass[12pt]{article}
\pdfoutput=1
\usepackage[numbers,sort&compress]{natbib}
\usepackage{graphicx}
\usepackage{amsmath}
\usepackage{amssymb}
\usepackage{hyperref}

\ifx\hypersetup\sadfkjashdfkxja\else
\hypersetup{pdftitle={Dressed Wilson Loops on S^2}}
\hypersetup{pdfauthor={Chrysostomos Kalousios, Donovan Young}}
\hypersetup{pdfsubject={}}
\hypersetup{pdfkeywords={}}
\hypersetup{plainpages=false}
\hypersetup{bookmarksnumbered=true}
\hypersetup{pdfstartview=FitH}
\hypersetup{pdfpagemode=UseNone}
\hypersetup{colorlinks=false}
\hypersetup{citebordercolor={.5 1 .5}}
\hypersetup{urlbordercolor={.5 1 1}}
\hypersetup{linkbordercolor={1 .7 .7}}
\fi

\newcommand{\ba}{\begin{aligned}}
\newcommand{\ea}{\end{aligned}}
\def\be{\begin{equation}}
\def\ee{\end{equation}}
\def\bsp{\be\begin{split}}
\def\la{\langle}
\def\ra{\rangle}

\def\L{\Lambda}

\def\a{\alpha}
\def\b{\beta}
\def\g{\gamma}

\def\d{\delta}
\def\e{\epsilon}
\def\m{\mu}

\def\n{\nu}
\def\s{\sigma}

\def\l{\lambda}
\def\t{\tau}

\def\vt{\vartheta}

\def\p{\partial}

\def\bR {\mathbb{R}}

\def\vp{\varphi}

\makeatletter

\newcommand{\Rmnum}[1]{\expandafter\@slowromancap\romannumeral #1@}
\makeatother

\newcommand{\beq}{\begin{equation}}
\newcommand{\eeq}{\end{equation}}
\newcommand{\bea}{\begin{eqnarray}}
\newcommand{\eea}{\end{eqnarray}}

\renewcommand{\title}[1]{\vbox{\center\LARGE{#1}}\vspace{5mm}}
\renewcommand{\author}[1]{\vbox{\center\large{#1}}\vspace{5mm}}
\newcommand{\address}[1]{\vbox{\center\em#1}}
\newcommand{\email}[1]{\vbox{\center\tt#1}\vspace{5mm}}

\newcommand{\Tr}{\mathrm{Tr}}

\newcommand{\Cset}{{\,\,{{{^{_{\pmb{\mid}}}}\kern-.47em{\mathrm C}}}}}

\newcommand{\comment}[1]{}
\begin{document}
\bibliographystyle{utphys}
\newpage
\setcounter{page}{1}
\pagenumbering{arabic}
\renewcommand{\thefootnote}{\arabic{footnote}}
\setcounter{footnote}{0}

\begin{titlepage}
\hfill {\tt HU-EP-11/19}\\
\title{\vspace{1.0in} {\bf Dressed Wilson Loops on $S^2$}}

\author{Chrysostomos Kalousios$^1$, Donovan Young$^2$}

\address{$^1$Humboldt-Universit\"at zu Berlin, Institut f\"ur Physik,\\
  Newtonstra\ss e 15, D-12489 Berlin, Germany,\\
$^2$Niels Bohr Institute, Blegdamsvej 17, DK-2100 Copenhagen, Denmark}

\email{$^1$ckalousi@physik.hu-berlin.de, $^2$dyoung@nbi.dk}

\abstract{We present a new, two-parameter family of string solutions
  corresponding to the holographic duals of specific 1/8-BPS Wilson
  loops on $S^2$ in ${\cal N}=4$ supersymmetric Yang-Mills theory. The
  solutions are obtained using the dressing method on the known
  longitude solution in the context of the auxiliary $\s$-model on
  $S^3$ put forth in arXiv:0905.0665[hep-th]. We verify that the
  regularized area of the world-sheets are consistent with
  expectations.}
\end{titlepage}

\section{Introduction}

The Maldacena-Wilson loop \cite{Maldacena:1998im,Rey:1998ik} has
proven to be of pervasive usefulness as an observable in the AdS/CFT
correspondence in particular \cite{Kristjansen:2010kg}, and in the
gauge-gravity duality in general. Examples of areas of application
include AdS/QCD \cite{CasalderreySolana:2011us}, scattering amplitudes
\cite{Alday:2010kn}, and Dp-brane theories
\cite{Brandhuber:1998bs,Agarwal:2009up,Chakraborty:2011ah}. In
Euclidean ${\cal N}=4$ supersymmetric Yang-Mills theory in four
dimensions, this non-local, gauge-invariant operator couples both to
the gauge field $A_\m(x)$ and to the six real scalar fields
$\Phi_I(x)$, through the trace of a path-ordered exponential
\be\label{wl}
W_R(C) = \frac{1}{\dim R(G)} \Tr_{R(G)} {\cal P} \exp \oint d\t
\Bigl( i\dot x^\m(\t) A_\m + |\dot x(\t)| \Theta^I(\t) \Phi_I\Bigr),
\ee
where the gauge group is denoted by $G$, $R$ denotes a representation
of $G$, and the path $C$ is defined by a closed contour
$\{x^\m(\t),\Theta^I(\t)\}$, where $\Theta^I(\t)\Theta^I(\t)=1$, and
so $\Theta^I(\t)$ defines a closed contour on $S^5$. The Wilson loop
defined in this manner enjoys local supersymmetry, and for contours
free of cusps, the scalar coupling removes would-be UV divergences at
coincident points along $C$. For particular choices of
$\{x^\m(\t),\Theta^I(\t)\}$, the supersymmetry can be enlarged to a
global symmetry of the operator, which leads to great simplifications
in the calculation of correlation functions, and in many cases, to
results valid for all values of the coupling constant $g_{\text{YM}}$
and the gauge group rank (e.g. $N$ for $G=SU(N)$, which will be the
gauge group of interest in what follows).

The AdS/CFT dictionary entry for this object is elegant and simple:
the contour $x^\m(\t)$ resides on the boundary of $AdS_5$ and provides
a boundary condition for open strings , $\Theta^I(\t)$ provides a
similar boundary condition on $S^5$. At large $N$ and $\l =
g^2_{\text{YM}}N$, and for $R$ of rank\footnote{For $R$ of rank ${\cal
  O}(N)$ the string is replaced by D3 and D5-branes
  \cite{Drukker:2005kx,Gomis:2006sb,Gomis:2006im,Hartnoll:2006is,Hartnoll:2006ib,Yamaguchi:2006tq}, for $R$ of rank ${\cal O}(N^2)$ the $AdS_5\times S^5$ background geometry is replaced by a back-reacted version \cite{Yamaguchi:2006te,Lunin:2006xr}.} ${\cal O}(1)$, the string
worldsheet is classical and describes a minimal surface in
$AdS_5\times S^5$ \cite{Drukker:1999zq}. The area\footnote{The area is
divergent and must be regularized by removing a term proportional to
the Wilson loop's perimeter.} of this minimal
surface is related to the logarithm of the expectation value of the
Wilson loop $\la W_R(C)\ra$. In this way we have replaced the problem
of the strong coupling behaviour of $W_R(C)$ by one of the most famous
problems of the calculus of variations, the Plateau problem, albeit in
a curved product space\footnote{For recent progress
  concerning Wilson loops with constant scalar coupling, see \cite{Ishizeki:2011bf}.}. Moving away from the classical limit,
$1/\sqrt{\l}$ corrections to  $\la W_R(C)\ra$ may be calculated via
semi-classical fluctuation determinants
\cite{Greensite:1998bp,Kinar:1999xu,Forste:1999qn,Drukker:2000ep,Kruczenski:2007cy,Roiban:2007dq,Kruczenski:2008zk,Chu:2009qt,Forini:2010ek,Faraggi:2011bb}.

The simplest Wilson loops (\ref{wl}) have
\be
\Theta^I(\t) = M^I_\m \,\frac{\dot x^\m(\t)}{|\dot x(\t)|},\qquad M^I_\m
M^I_\n = \d_{\m\n},
\ee
and were discovered by Zarembo in \cite{Zarembo:2002an}. If the curve
$x^\m(\t)$ lies in $\bR^n$, these Wilson loops are $(1/2)^n$-BPS. They
have trivial expectation value $\la W \ra = 1$, to all orders of
perturbation theory
\cite{Guralnik:2003di,Guralnik:2004yc,Kapustin:2006pk}. The
next-to-simplest Wilson loop (\ref{wl}) is the 1/2-BPS circle with contour
$\{(\cos\t,\sin\t,0,0),\Theta^I=\text{const.}\}$ whose dynamics are captured
by a Hermitian matrix model
\cite{Erickson:2000af,Drukker:2000rr,Pestun:2007rz}, exact for all
values of $N$ and $\l$. Recently, these two examples of Wilson loops
were shown to arise from a larger class of generically 1/16-BPS Wilson
loops with $x^\m(\t) \subset S^3$, and with scalar coupling given by \cite{Drukker:2006ga,Drukker:2007dw,Drukker:2007yx,Drukker:2007qr}
\be
\Theta^I(\t) = \frac{1}{|\dot x|} \s^i_{\m\n} x^\m \dot x^\n M_i^I,\qquad M_i^I
M_j^I = \d_{ij},
\ee
where the tensor $\s^i_{\m\n}$ is defined via the projection of the
Lorentz generators in the anti-chiral spinor representation ($\g_{\m\n}$) onto the
Pauli matrices $\t^i$
\be
\frac{1}{2}(1-\g^5) \g_{\m\n} = i \s^i_{\m\n}  \t_i.
\ee
In the special case when $x^4(\t)=0$, the contour $\vec x(\t)$ resides
on a great $S^2\subset S^3$, the Wilson loops are 1/8-BPS, and
$|\dot{\vec x}|\Theta^I=(\vec x \times \dot{ \vec
x},0,0,0)$. Incredibly, these 1/8-BPS loops on $S^2$ appear to be
captured exactly by the zero-instanton sector of pure Yang-Mills in
two-dimensions
\cite{Bassetto:2008yf,Young:2008ed,Bassetto:2009rt,Bassetto:2009ms,Giombi:2009ds,Pestun:2009nn,Giombi:2009ms}\footnote{An
original disagreement for Wilson loop correlators presented in
\cite{Young:2008ed} has since been retracted; the numerical analysis
presented in \cite{Bassetto:2009ms} supports the two-dimensional
Yang-Mills conjecture.}, and therefore by a matrix model. The result
for single Wilson loop VEV's is\footnote{$L_n^m$ is the Laguerre
polynomial $L_n^m(x)=1/n!\exp[x]x^{-m}(d/dx)^n (\exp[-x]x^{n+m})$.}
\be \label{L}
\la W \ra = \frac{1}{N} L_{N-1}^1\left(-g_{\text{YM}}^2 \frac{A_1
  A_2}{A^2}\right)\exp \left(\frac{g_{\text{YM}}^2}{2} \frac{A_1 A_2}{A^2}\right),
\ee
where $A_1$ is the area on $S^2$ enclosed by $\vec x(\t)$ and $A_2 =
A-A_1$, where $A$ is the total sphere area. When $\vec x(\t) =
(\cos\t,\sin\t,0)$ one recovers the 1/2-BPS circle and the associated
result from the Hermitian matrix model. The limit of a latitude
shrinking to zero size at the north pole gives a Zarembo circle, and
$\la W\ra = 1$.

At strong coupling and large-$N$, the 1/8-BPS Wilson loops on $S^2$
enjoy a description which is a generalization of the calibrated
surfaces technique originally applied to the Zarembo loops at strong
coupling in \cite{Dymarsky:2006ve}. In particular, the problem of
finding classical string solutions of minimal area which end on the
1/8-BPS contours can be reduced to a sigma-model on $S^3$
\cite{Giombi:2009ms}. The procedure is non-trivial however, and not
every solution of the sigma-model provides a Wilson loop. Indeed,
there were originally only two solutions known, and these were obtained without
recourse to the sigma-model: the latitude and the loop formed by two
longitudes (i.e. an ``orange-wedge'') \cite{Drukker:2007qr}. The sigma-model
allowed a coincident latitude-latitude solution, and an approximate
perturbed latitude solution to be found \cite{Giombi:2009ms}. In the present paper, we will
use the dressing method on the longitude solution to find new
solutions, whose boundary curves are non-trivial shapes on $S^2$. In
so-doing we can verify that the regularized area of the worldsheet is
in accordance with (\ref{L}).

This paper is organized as follows. We begin with a review of the
pseudo-holomorphicity equations in section \ref{sec:pseudo}. In
section \ref{sec:dress} we review the dressing method and show how the
longitude solution is obtained by dressing. We continue in section
\ref{sec:petal} with a presentation of new solutions obtained by
dressing the longitude solution. We conclude in section \ref{sec:disc}
with a discussion.

\section{Pseudo-holomorphicity equations and sigma-model on $S^3$}
\label{sec:pseudo}

The 1/8-BPS Wilson loops on $S^2$ couple to
three of the six scalar fields of ${\cal N}=4$ SYM $\vec \Phi$
\be
W = \frac{1}{N} \Tr \,{\cal P} \exp \oint d\t \left( i \dot x^\mu A_\mu +
(\vec x \times \dot{\vec x})\cdot \vec\Phi \right),
\ee
where
\be
x^\mu = (\vec x, 0), \qquad \vec x^2 =1.
\ee
The string duals of
these Wilson loops are contained in an $AdS_4 \times S^2$ subspace of
$AdS_5\times S^5$. We write the metric of this subspace as
\be
ds^2 = \frac{1}{z^2} dx^i dx^i + z^2 dy^i dy^i,\qquad z^2 = \frac{1}{y^iy^i},
\ee
where $i=1,\ldots,3$, so that $\vec\theta \equiv z\vec y$ are
embedding coordinates for $S^2$.

The worldsheets defined by $\{\vec x(\s,\t),\,\vec y(\s,\t),\,z(\s,\t)\}$
obey first order differential equations (``pseudo-holomorphicity
equations'') arising from supersymmetry and some further
non-differential constraints \cite{Giombi:2009ms}. These are as
follows\footnote{We take $\p_0 = \p_\s$ and $\p_1 = \p_\tau$, while
  $\e_{01} = -\e_{10}=1$.}
\bsp\label{cons}
\vec x^2 + z^2 = 1,\qquad \vec x \cdot \vec y = C=\text{const.},\qquad
z^2 \p_\a (\vec x \times \vec y ) = \e_{\a\b} \p_\b \vec x.
\end{split}
\ee
It is these relations which allow the problem of finding solutions to
be reduced to an auxiliary sigma model on $S^3$, supplemented with
non-trivial added constraints. One defines the following 4-vector
\be\label{xidef}
\xi^A = (\vec \xi,\xi^4),\qquad \vec \xi = z \vec y \times
\vec x,\qquad
\xi^4 = \sqrt{1+C^2} z.
\ee
Using the first two relations of (\ref{cons}), one may show that
$\xi^A \xi^A =1$, and $\xi^A$ is therefore contained in an
$S^3$. Further, by operating $\p^2$ upon $\xi^A \xi^A$, one obtains
\be
\p^2 \xi^A + \xi^A \p_\a \xi^B \p_\a \xi^B = 0,
\ee
which are the equations of motion of the sigma model
\be
S=\frac{1}{2} \int d^2\s \left( \p_\a \xi^A \p_\a \xi^A + \L ( \xi^A
\xi^A -1) \right).
\ee
The last relation in (\ref{cons}) allows one to integrate a solution
to the sigma-model in order to obtain $\vec x$
\be\label{intx}
\p_\a \vec x = \frac{1}{\sqrt{1+C^2}} \e_{\a\b} \left( \xi^4 \p_\b
\vec \xi - \vec \xi \p_\b \xi^4 \right).
\ee
The non-differential constraints can be used to show that
\be\label{y}
\vec y = \frac{1}{1-z^2} \left( \frac{\vec x \times \vec \xi}{z} + C
\vec x \right),
\ee
which gives $\vec y$ once $\vec x$ is known, but only if one ensures
that
\be\label{addcons}
\vec \xi \cdot \vec x=0, \qquad \vec x^2 + \frac{1}{\vec y^2} = 1,
\ee
which are necessary in order to be consistent with
(\ref{xidef}). These additional constraints greatly constrain the number of solutions to the
sigma-model which actually correspond to Wilson loop surfaces. The
boundary of the string needs to end on the boundary of $AdS_5$ along
the Wilson loop contour. This is ensured by the following boundary
conditions on $\xi^A$
\be
\xi^4 |_\p = 0,\qquad \vec \xi|_\p
= \left. \frac{\dot{\vec x}}{|\dot{\vec{x}}|}\right|_\p,
\ee
where the dot denotes the derivative along the boundary curve.

The regularized area of the worldsheet has a simplified form owing to
the pseudo-holomorphicity equations \cite{Giombi:2009ms}
\be
S_{\text{reg.}} = \frac{\sqrt{\l}}{4\pi} \int d^2\s \left(\p_a \vec
\theta \cdot \p^a \vec \theta + \frac{1}{z} \nabla^2 z \right),
\ee
and was shown to be invariant under area-preserving diffeomorphisms,
from which one can use known solutions to fix the answer to the result
expected from (\ref{L}), i.e.
\be\label{Sreg}
S_{\text{reg.}} = -\frac{\sqrt{\l}}{2\pi} \sqrt{A_1 A_2},
\ee
where we remind the reader that $A_1$ is the area on $S^2$ enclosed by
the Wilson loop, and $A_2$ is the conjugate area. A corollary of that
same analysis showed that
\be\label{Cexp}
C = \pm \frac{A_2-A_1}{2\sqrt{A_1 A_2}}.
\ee
where the $\pm$ refer to the stable/unstable conjugate wrappings of
the $S^2 \subset S^5$, see \cite{Giombi:2009ms} for a discussion. In
the body of the paper we will always give the stable solution.

\begin{figure}
\begin{center}
\includegraphics[height=1.5in]{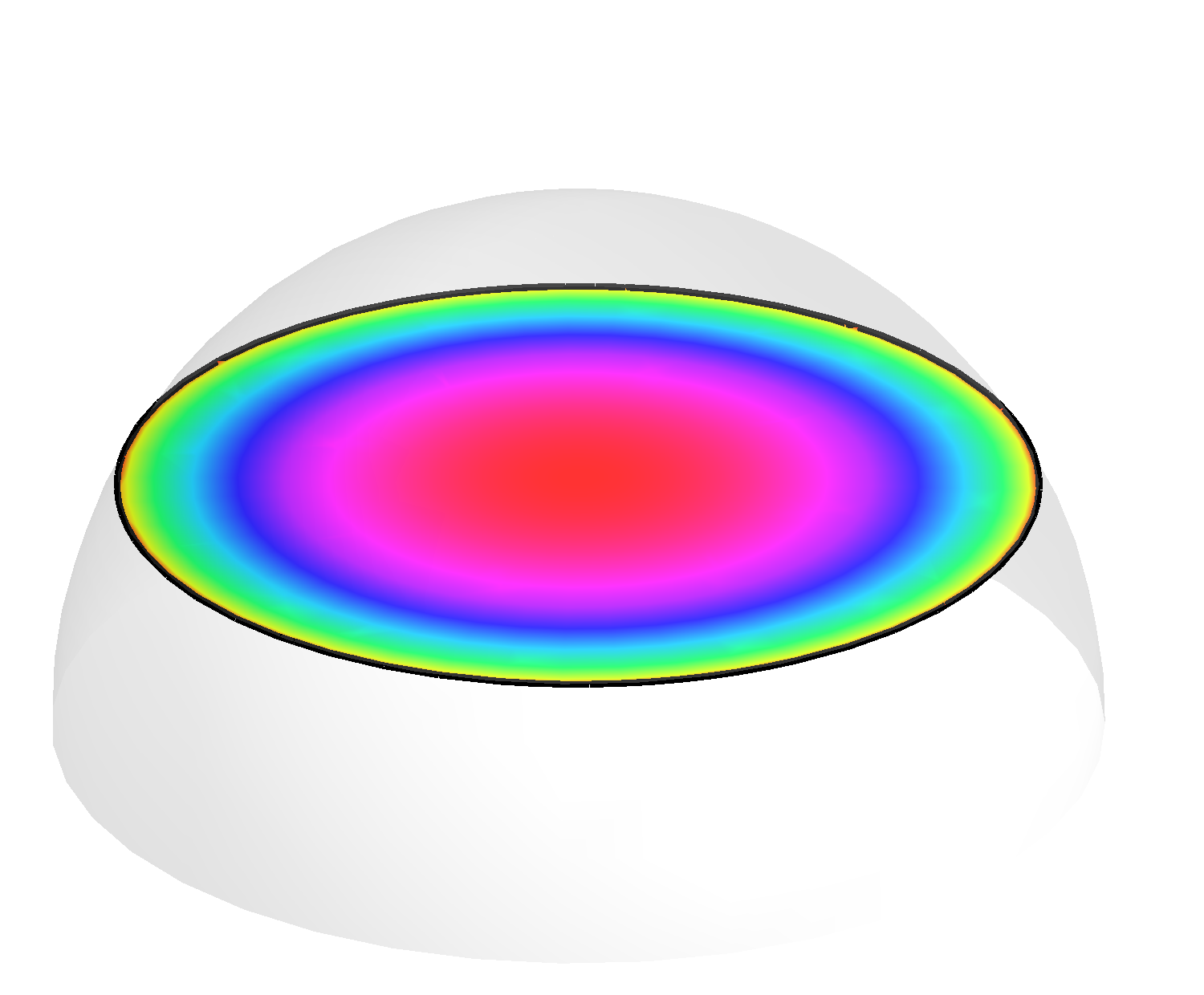}\hspace{0.75cm}
\includegraphics[height=1.5in]{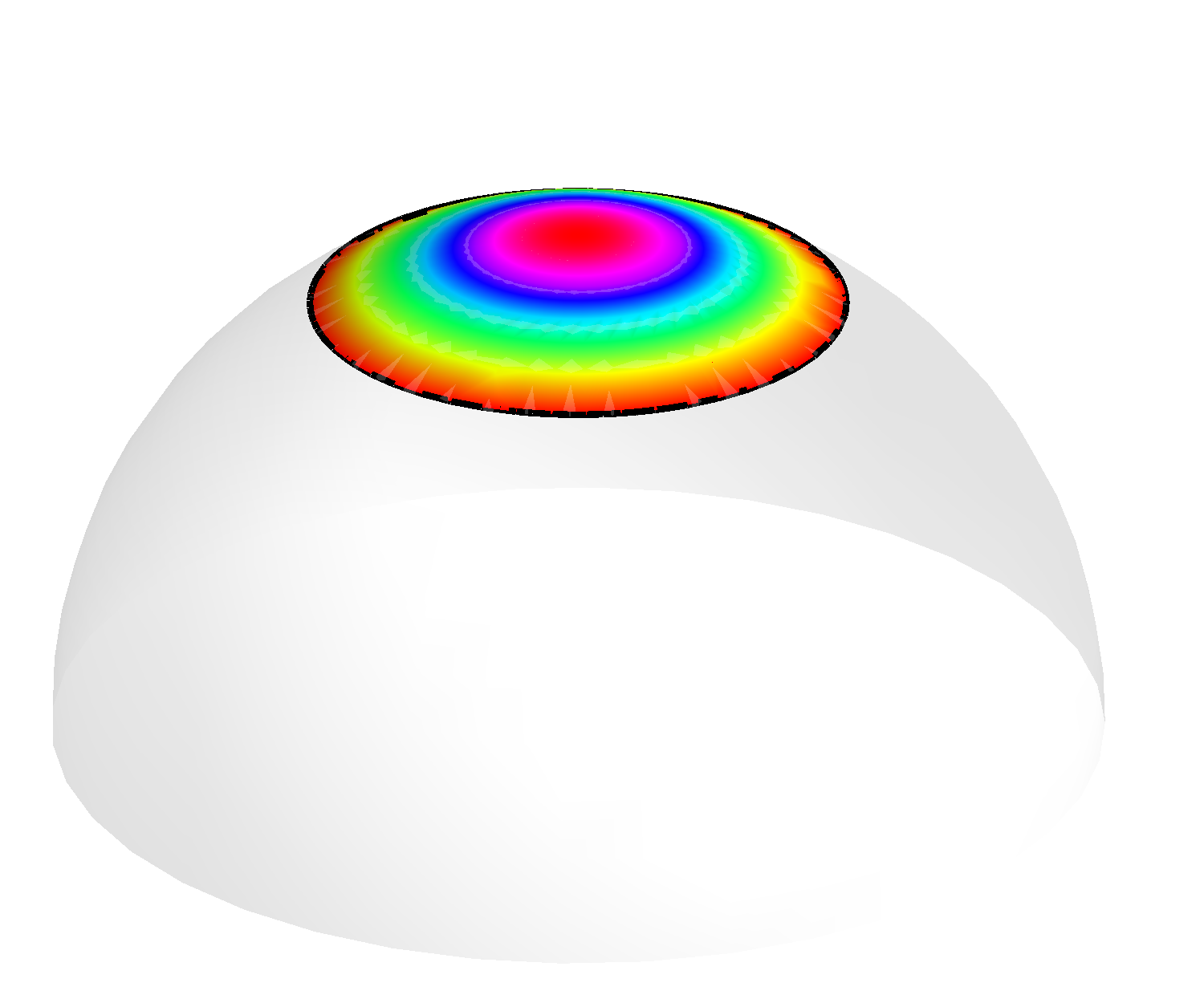}\hspace{0.75cm}
\includegraphics[height=1.5in]{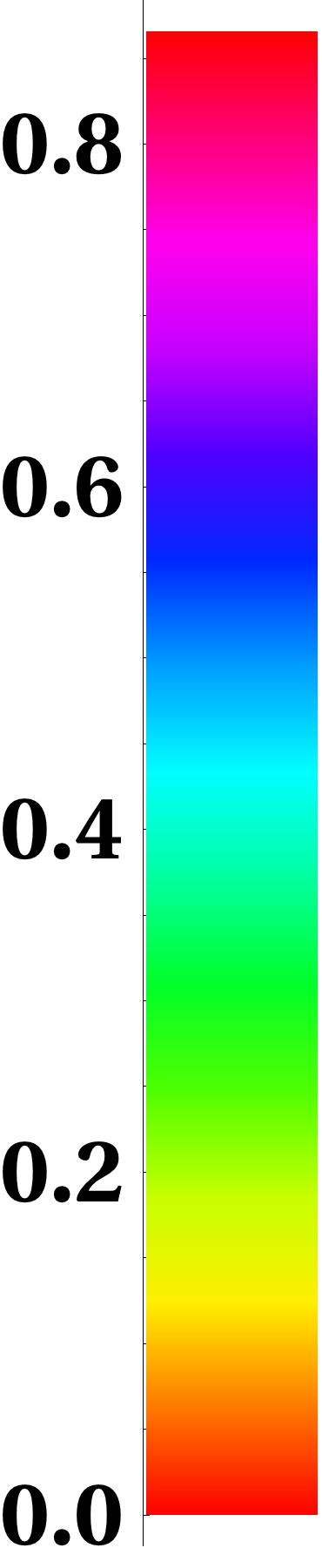}
\end{center}
\caption{The latitude solution, from left to right: the surface $\vec
x(\s,\t)$ - its boundary (black curve) lies on the unit $S^2$ (shown
in gray) $\subset\p AdS_5$; the surface $\vec\theta(\s,\t)$ - lying entirely on the unit
$S^2\subset S^5$. The two surfaces are coloured according to the value
of the $AdS_5$ $z$-coordinate, i.e. by $z(\s,\t)$, whose scale is given
in the last plot.}
\label{fig:lat}
\end{figure}

There are two canonical solutions known from the literature. They are
the previously mentioned latitude and longitude solutions. The
latitude solution is given by
\bsp
&\vec x = \sin\theta_0
\left(\frac{\cos\t}{\cosh\s},\frac{\sin\t}{\cosh\s},\cot\theta_0\right),\quad
z = \sin\theta_0\tanh\s, \quad C = \pm \cot\theta_0,\\
&\xi^A =
\left(-\frac{\sin\t}{\cosh\s},\frac{\cos\t}{\cosh\s},0,\tanh\s\right),\quad
\s\in[0,\infty],\quad \t\in[0,2\pi],
\end{split}
\ee
and $\vec y$ is given by (\ref{y}). This solution is shown in figure
\ref{fig:lat}. The boundary curve $\vec x|_{z=0}$ is a latitude at
polar angle $\theta_0$.

\begin{figure}
\begin{center}
\includegraphics[height=1.75in]{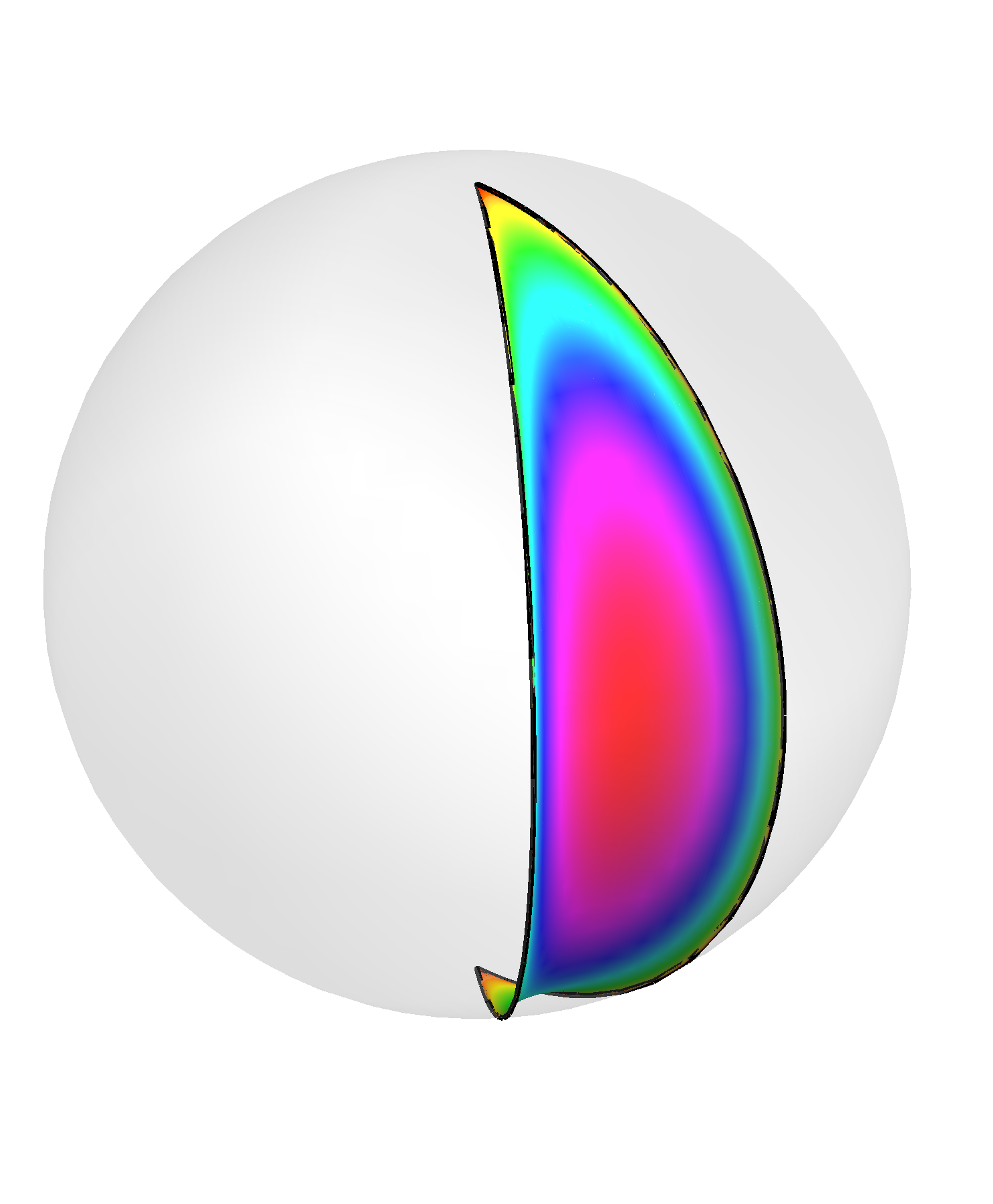}\hspace{1.75cm}
\includegraphics[height=1.75in]{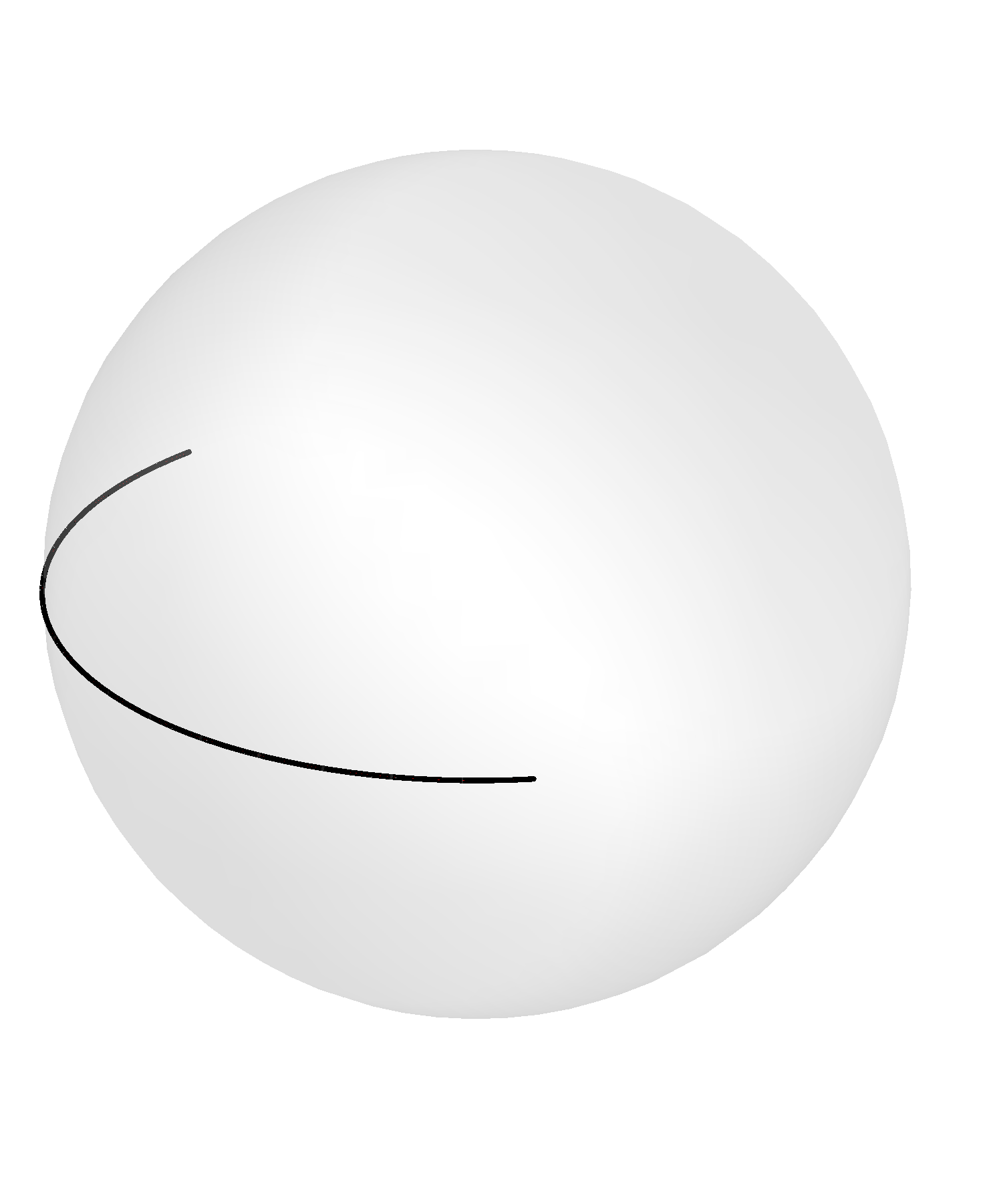}\hspace{1.75cm}
\includegraphics[height=1.75in]{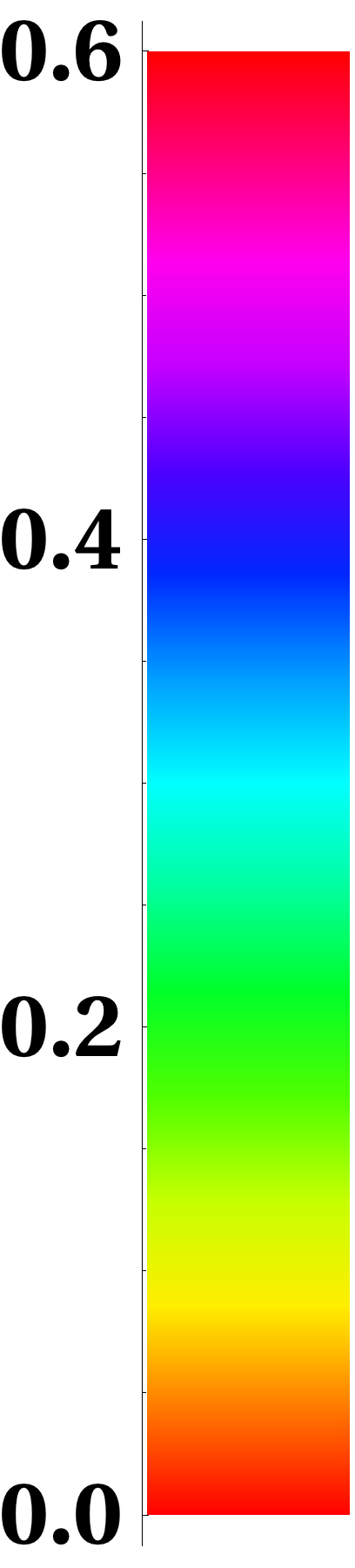}
\end{center}
\caption{The longitudes solution, from left to right: the surface $\vec
x(\s,\t)$ - its boundary (black curve) lies on the unit $S^2$ (shown
in gray) $\subset\p AdS_5$; the surface $\vec\theta(\s,\t)$ - lying entirely on the unit
$S^2\subset S^5$. The two surfaces are coloured according to the value
of the $AdS_5$ $z$-coordinate, i.e. by $z(\s,\t)$, whose scale is given
in the last plot.}
\label{fig:long}
\end{figure}

The longitude solution is given by
\bsp\label{long}
&\vec x = \left(\frac{a\sin a\s \sin\s + \cos a\s
  \cos\s}{\cosh\sqrt{1-a^2}\t},
\frac{a\cos a\s \sin\s - \sin a\s
  \cos\s}{\cosh\sqrt{1-a^2}\t}, -\tanh\sqrt{1-a^2}\t \right),\\
&z = \frac{\sqrt{1-a^2}\sin\s}{\cosh\sqrt{1-a^2}\t},\quad
C = \frac{a}{\sqrt{1-a^2}},\quad
\s\in[0,\pi],\quad \t\in[-\infty, \infty],\\
&\xi^A =  \tanh \sqrt{1-a^2}\t\left(-\cos a\s, \sin a\s ,
-\frac{\cos\s}{\sinh\sqrt{1-a^2}\t}, \frac{\sin\s}{\sinh
  \sqrt{1-a^2}\t}\right),
\end{split}
\ee
which is shown in figure \ref{fig:long}. The boundary curve $\vec
x|_{z=0}$ is given by two longitudes with opening angle $(1-a)\pi$. We will see that the
new solutions of section \ref{sec:petal} degenerate to this
solution in a particular limit. The longitude solution may be obtained by dressing the
``vacuum'' solution $\xi^A = (\sin\t,\cos\t,0,0)$ once, as we will
show in the next section. The new solutions presented in section
\ref{sec:petal} are obtained by dressing the vacuum
twice, i.e. by dressing the longitude solution.

\section{Dressing method}
\label{sec:dress}

We use the dressing method \cite{Zakharov:1973pp} for the $SU(2)$ sigma model to construct
strings that live in $S^3$ and have Euclidean worldsheet.  Let $\xi^A$
be the spacetime components of the string.  We consider the element
\be g=\begin{pmatrix} \xi^1+i \xi^2 & -i(\xi^3+i \xi^4) \\ -i(\xi^3-i \xi^4) &
\xi^1-i \xi^2
  \end{pmatrix} \in SU(2)
\ee
and as vacuum we take
\be
\xi^A=(\sin \tau,~\cos\tau,~0,~0)^{\rm T}.
\ee
Going to lightcone coordinates $z_\pm=(\sigma \pm i \tau)/2$ we seek a solution to the system of equations
\be
\partial_\pm\Psi(\lambda)=\frac{\partial_\pm g g^{-1}}{1\pm i\lambda} \Psi(\lambda)
\ee
subject to the initial condition $\Psi(0)=g$ and the coset constraint $\Psi(\bar{\lambda})^\dagger \Psi(\lambda)=I$.  We find
\be
\Psi(\lambda)=
\begin{pmatrix}
 i e^{-i f(\lambda)} & 0 \cr
 0 & -i e^{i f(\lambda)}
\end{pmatrix},\quad f(\lambda) = \frac{\lambda \sigma + \tau}{1+\lambda^2}.
\ee
The general $N$-soliton solution for the $SU(n)$ sigma model has been constructed in \cite{Kalousios:2010ne} (see also \cite{Harnad:1982cf,Spradlin:2006wk,Kalousios:2008gz}).  Here we are interested in the special case $n=2$ and we are focusing only on 1- and 2-soliton solutions, which can be expressed respectively as
\be\label{1magnon}
g_1 =\sqrt{ \frac{\lambda_1}{\bar{\lambda}_1}} \frac{(\alpha_{11} I_{2\times 2} + h_1 h_1^\dagger) g_0}{\alpha_{11}}
\ee
and
\be\label{2magnons}
g_2 =\sqrt{\frac{\lambda_1 \lambda_2}{\bar{\lambda}_1 \bar{\lambda}_2}} \frac{\left( (\a_{11}\a_{22}-\a_{12}\a_{21})I_{2\times 2}+\a_{22} h_1 h_1^\dagger+\a_{11} h_2  h_2^\dagger-\a_{12} h_1  h_2^\dagger-\a_{21} h_2  h_1^\dagger \right) g_0}{\a_{11}\a_{22}-\a_{12}\a_{21}}.
\ee
In the above
\be
h_i=\Psi(\bar{\lambda}_i) e_i, \quad \beta_{ij}=h_i^\dagger h_j, \quad \a_{ij}=-\frac{\l_i \beta_{ij}}{\l_i -\bar{\l}_j}.
\ee
The arbitrary complex vectors $e_i$ are called polarization vectors
and the complex numbers $\l_i$ are the spectral parameters of the
problem. Once a dressed solution is found, one must then integrate
(\ref{intx}) and impose the constraints (\ref{cons}) and
(\ref{addcons}) to find a Wilson loop solution. In general it is not
possible to satisfy these constraints. We have found a two-parameter
family of solutions which do lead to Wilson loops, and these are
presented in section \ref{sec:petal}.

\subsection{Longitude from dressing}

We can reproduce the known longitude solution \eqref{long} from
\eqref{1magnon}.  We choose the polarization vector to be
$e_1=(1+i,\,1+i)^{\rm T}$ and the spectral parameter
$\lambda_1=\sqrt{\frac{1+a}{1-a}} \,i$, $|a|<1$.  Then we can easily
see that the sigma model solution agrees with \eqref{long}.

\section{Petal solutions}
\label{sec:petal}

The twice-dressed vacuum (\ref{2magnons}) leads to a new two-parameter
family of string worldsheets dual to 1/8-BPS Wilson loops. We take
the spectral parameters to have only imaginary parts, namely we take
$\l_1 = b\,i,~\l_2=i/b$. We choose the polarization vectors to be
$e_1=(i,~-i)^{\rm T},~e_2=(a\, i,~i)^{\rm T}$. The conditions
(\ref{cons}) and (\ref{addcons}) fix $a$ and the solution\footnote{After obtaining the
solution from the dressing method, we flip the sign of $\vec\xi$,
which is a symmetry of the sigma-model on $S^3$. This fixes
conventions to the standard ones, where $C>0$ corresponds to the
stable string worldsheet.} is as follows
\bsp
&\vec x=\left(\frac{N_1}{D},\frac{N_2}{D},\frac{N_3}{D} + 1\right),\quad
\xi^A = (\vec\xi,\xi^4)=\frac{b\sqrt{1+C^2}}{D} (Z_1,Z_2,Z_3,Z_4),\\
&z=\frac{\xi^4}{\sqrt{1+C^2}},\quad a = -\frac{1+b^2 -
  2b\sqrt{1+C^2}}{1+b^2 + 2b\sqrt{1+C^2}},\quad \s \in
   [\s_0,\infty],\quad\t\in\left[0,\pi \frac{b^2-1}{b^2+1}\right],
\end{split}
\ee
where $\vec y$ is given by (\ref{y}), $\s_0$ is given by (\ref{s0}), and where
\bsp
&N_1 = -(1+a)(b^4-1)e^{2b\s/(b^2-1)} \Bigl( a + e^{4b\s/(b^2-1)} \Bigr)
\Bigl(\sin\frac{2b^2\t}{1-b^2} + b^2\sin\frac{2\t}{b^2-1}\Bigr),\\
&N_2 = (1+a)(b^4-1)e^{2b\s/(b^2-1)} \Bigl( a + e^{4b\s/(b^2-1)} \Bigr)
\Bigl(\cos\frac{2b^2\t}{1-b^2} + b^2\cos\frac{2\t}{b^2-1}\Bigr),\\
&N_3 = -(1-a^2)(1-b^2)^2(1+b^2)e^{4b\s/(b^2-1)},\\
&D = b\sqrt{1+C^2} \Biggl[ (1+b^2)^2
\left(a^2+e^{8b\s/(b^2-1)}\right)\\
&\qquad + e^{4b\s/(b^2-1)} \left((1+a^2)(1-b^2)^2+8a b^2
\cos\, 2\frac{b^2+1}{b^2-1}\t \right)\Biggr],
\end{split}
\ee
and where
\bsp
&Z_1= (1+b^2)^2\left(a^2+e^{8b\s/(b^2-1)}\right)\sin\t\\
&\qquad- e^{4b\s/(b^2-1)} \left((1+a^2)(1-b^2)^2\sin\t + 4a \left(
b^4\sin\,\frac{b^2+3}{b^2-1}\t +
\sin\,\frac{3b^2+1}{1-b^2}\t\right)\right),\\
&Z_2= (1+b^2)^2\left(a^2+e^{8b\s/(b^2-1)}\right)\cos\t\\
&\qquad- e^{4b\s/(b^2-1)} \left((1+a^2)(1-b^2)^2\cos\t - 4a \left(
b^4\cos\,\frac{b^2+3}{b^2-1}\t +
\cos\,\frac{3b^2+1}{1-b^2}\t\right)\right),\\
&Z_3= -2(1-a)(b^4-1)
e^{2b\s/(b^2-1)}\left(a+e^{4b\s/(b^2-1)}\right)\cos\,\frac{b^2+1}{b^2-1}\t,\\
&Z_4= 2(1+a)(b^4-1)
e^{2b\s/(b^2-1)}\left(-a+e^{4b\s/(b^2-1)}\right)\sin\,\frac{b^2+1}{b^2-1}\t.
\end{split}
\ee
The boundary curve $\vec x|_{z=0}$ consists of two longitudes
emanating from the north pole, given by $\t = 0$, $\pi
(b^2-1)/(b^2+1)$ and $\s\in[\s_0,\infty]$, and a curve connecting
their endpoints, given by $\s=\s_0$ and $\t\in[0,\,\pi
(b^2-1)/(b^2+1)]$, where
\be\label{s0}
\s_0 = \frac{(b^2-1)}{4b} \log a.
\ee
We note that $b$ and $C$ must be chosen so that $a > 0$, i.e.
\be
\sqrt{1+C^2}-C < b < \sqrt{1+C^2}+C.
\ee
At the special values $b =  \sqrt{1+C^2}\pm C$ the
solution degenerates to the longitude solution (\ref{long}).

\begin{figure}
\begin{center}
\includegraphics[height=1.5in]{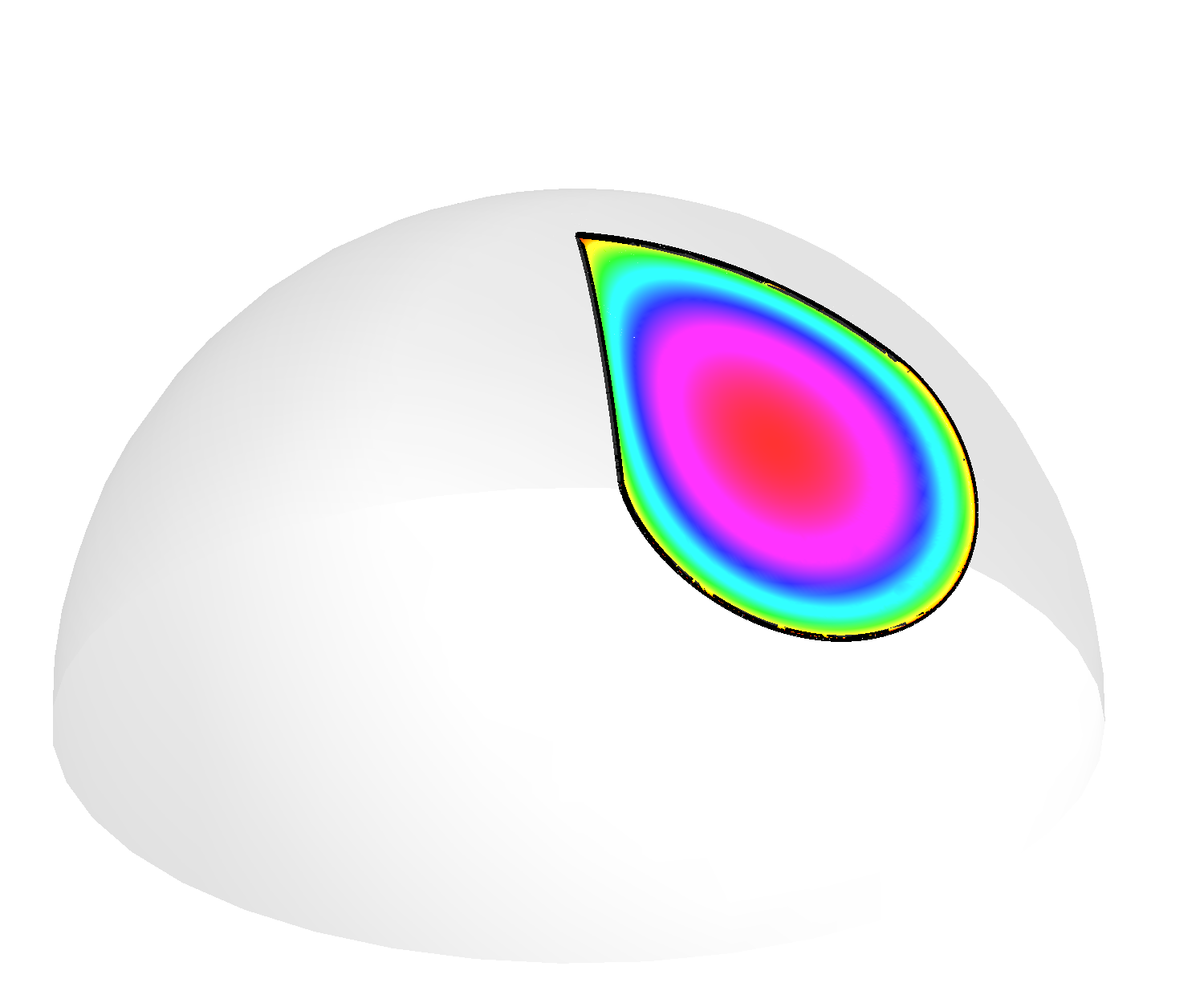}\hspace{0.5cm}
\includegraphics[height=1.15in]{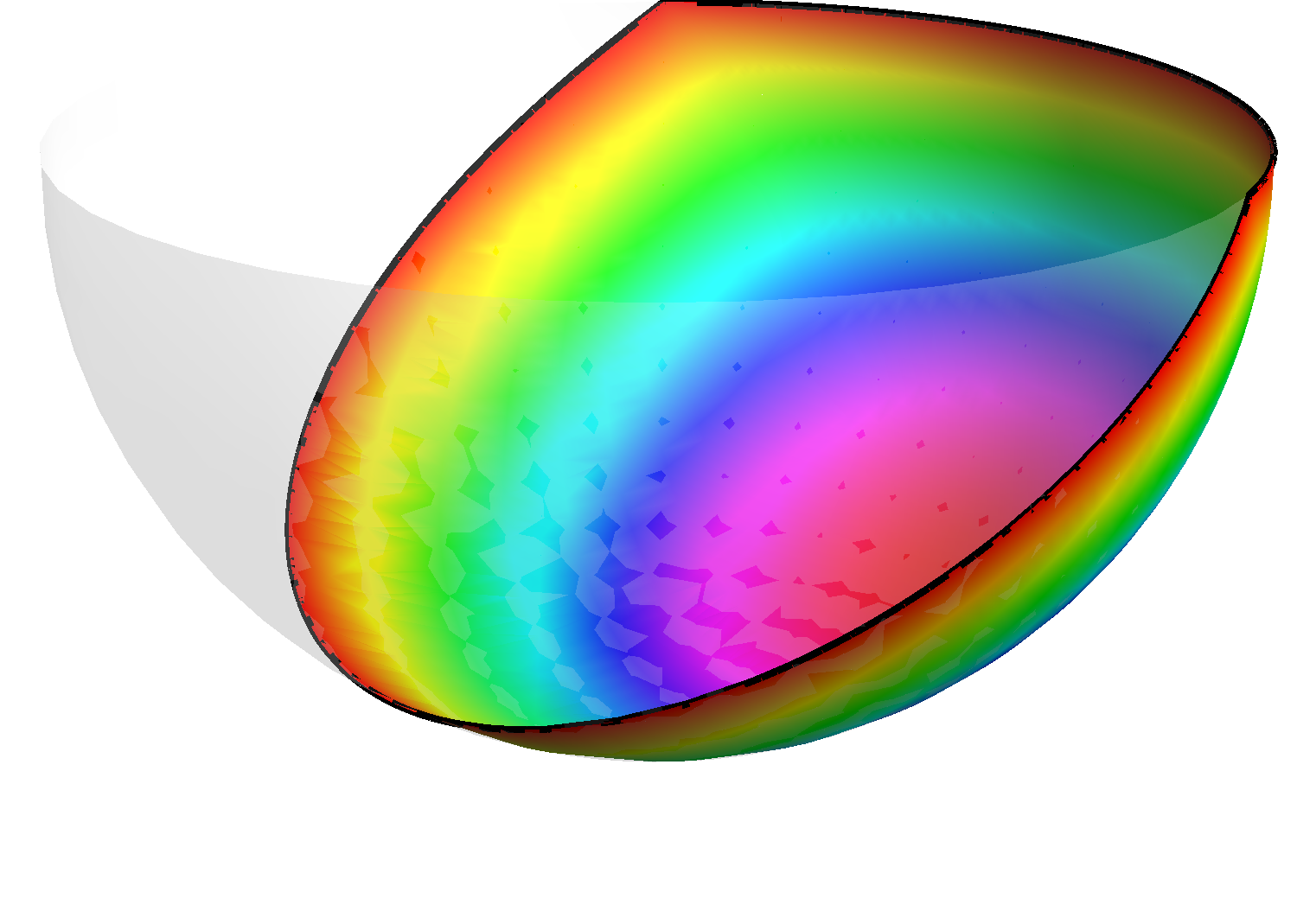}\hspace{0.5cm}
\includegraphics[height=1.5in]{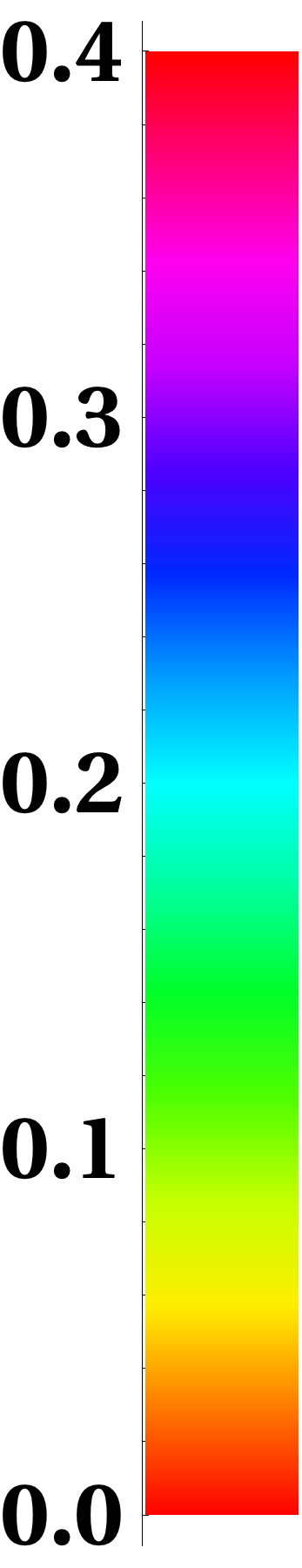}
\end{center}
\caption{The ``petal'' solution for $b=2$ and $C=\sqrt{21}/2$, from left to right: the surface $\vec
x(\s,\t)$ - its boundary (black curve) lies on the unit $S^2$ (shown
in gray) $\subset\p AdS_5$; the surface $\vec\theta(\s,\t)$ - lying entirely on the unit
$S^2\subset S^5$. The two surfaces are coloured according to the value
of the $AdS_5$ $z$-coordinate, i.e. by $z(\s,\t)$, whose scale is given
in the last plot.}
\label{fig:petal}
\end{figure}

\begin{figure}
\begin{center}
\includegraphics[height=1.25in]{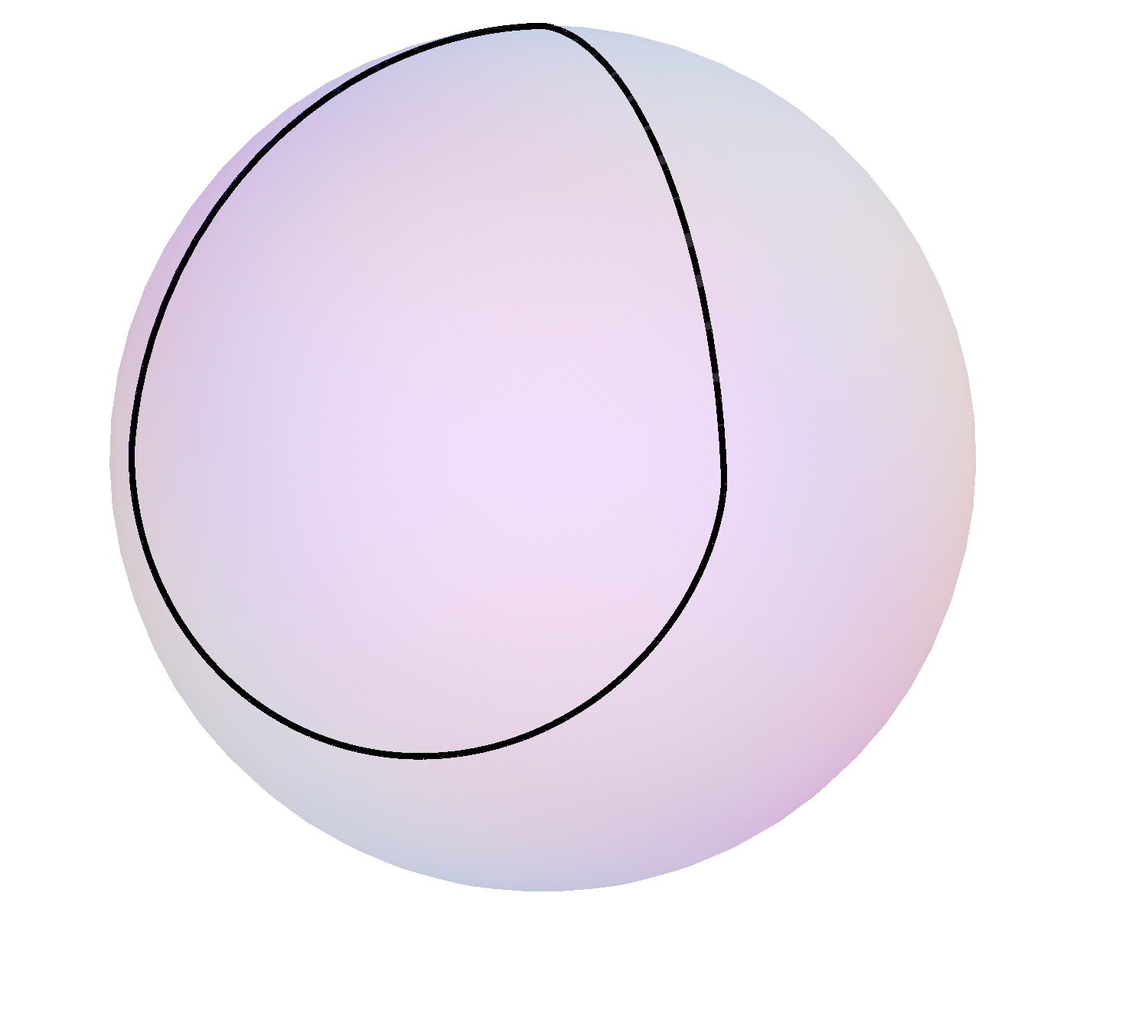} 
\includegraphics[height=1.25in]{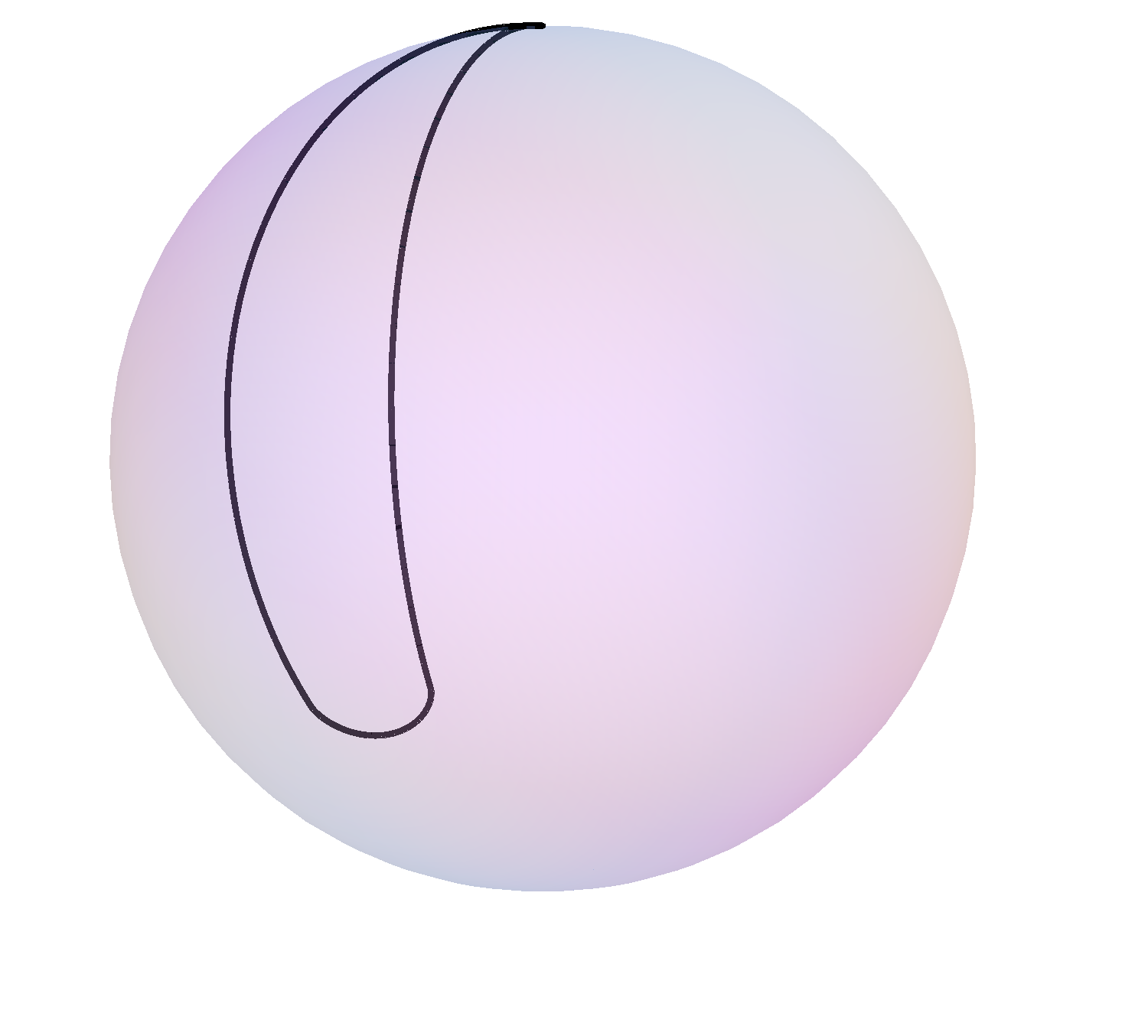} 
\includegraphics[height=1.25in]{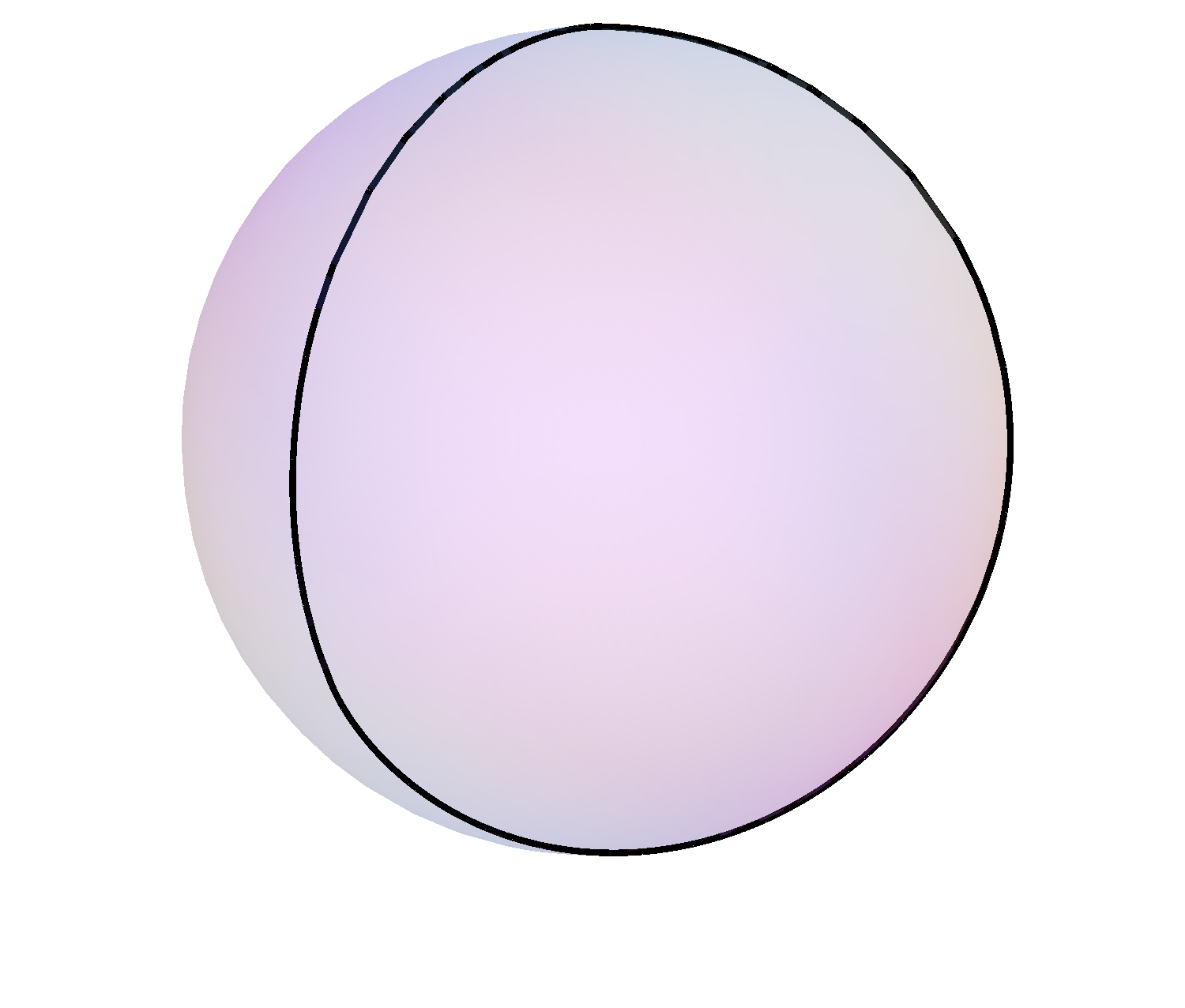}\\
\includegraphics[height=1.25in]{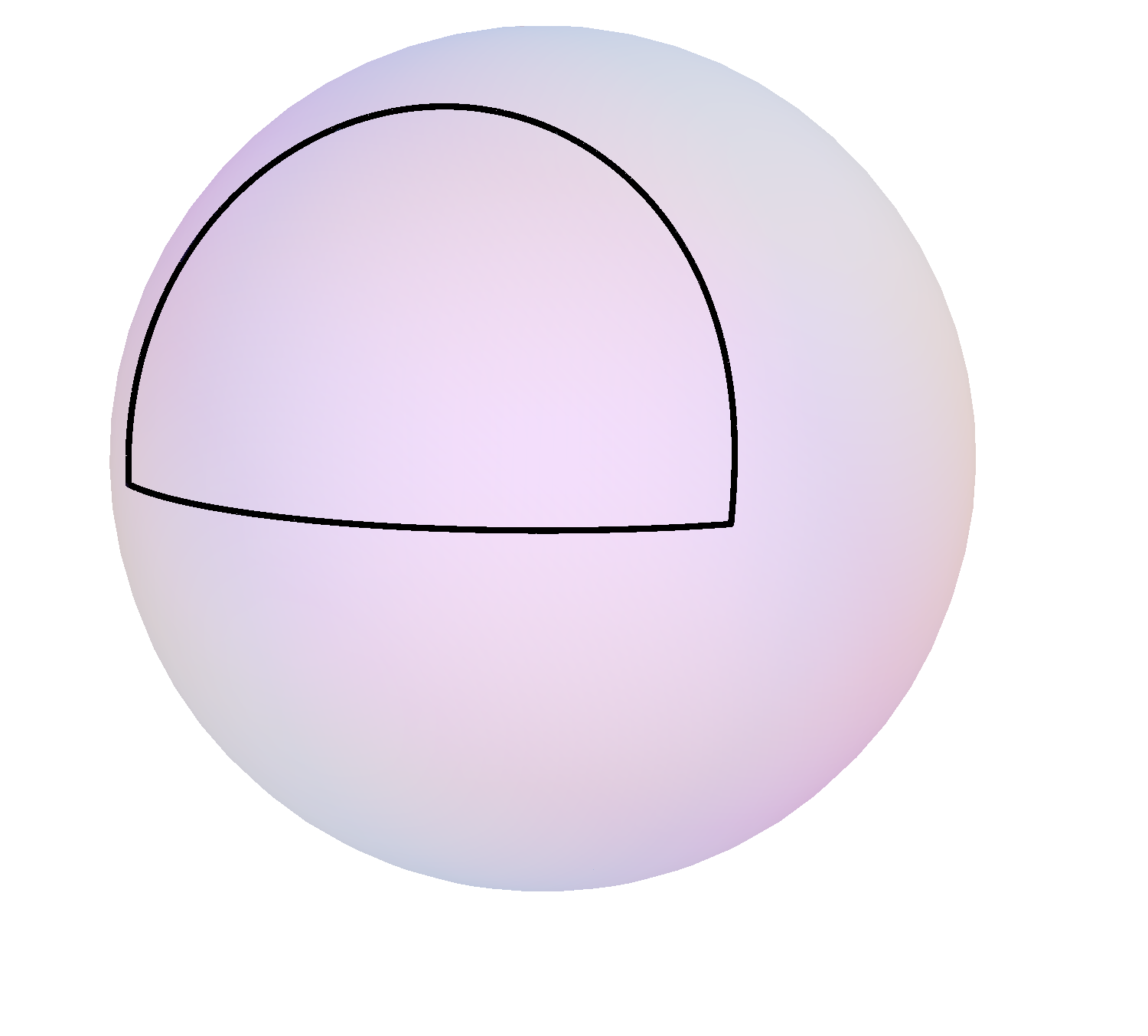} 
\includegraphics[height=1.25in]{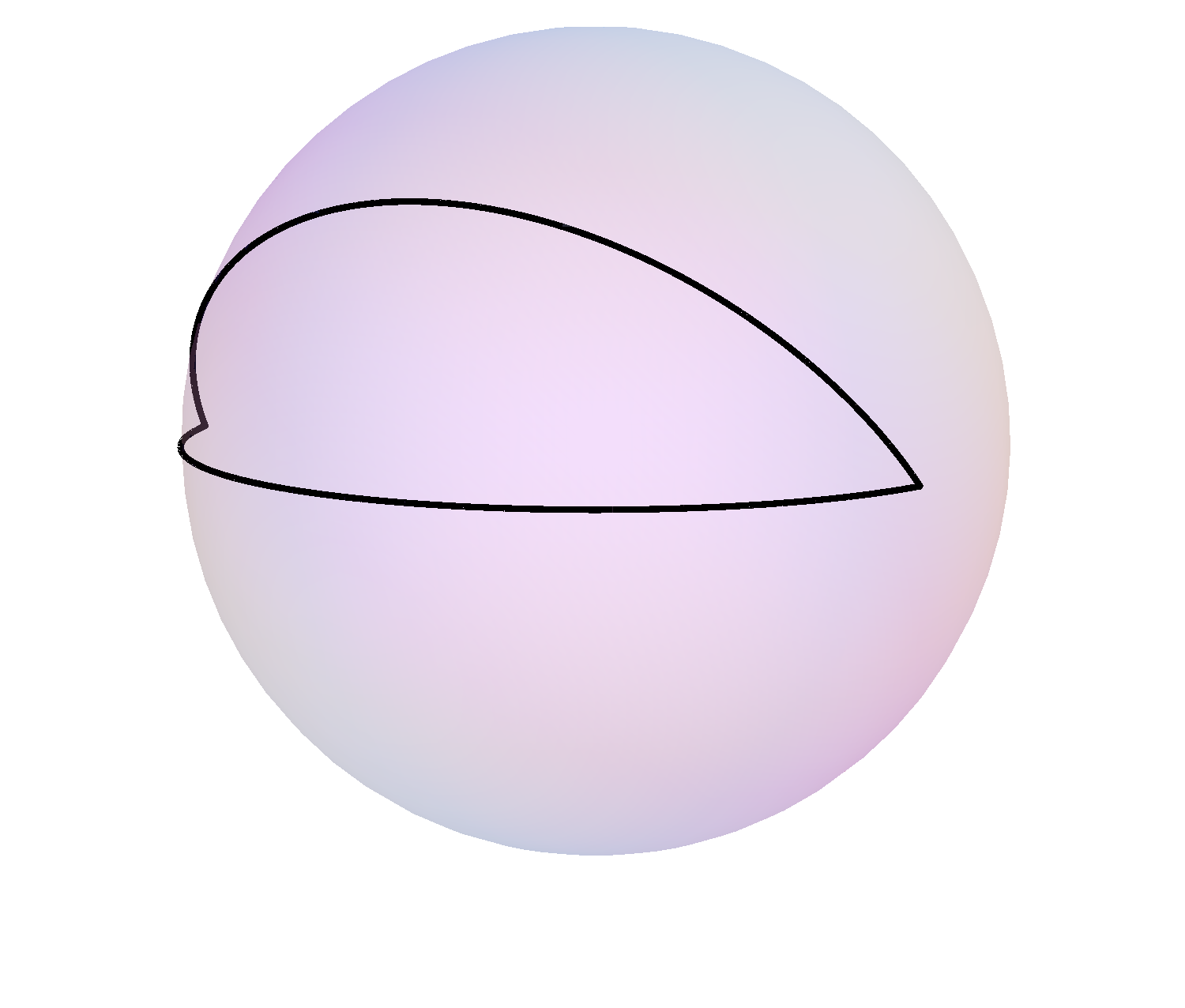} 
\includegraphics[height=1.25in]{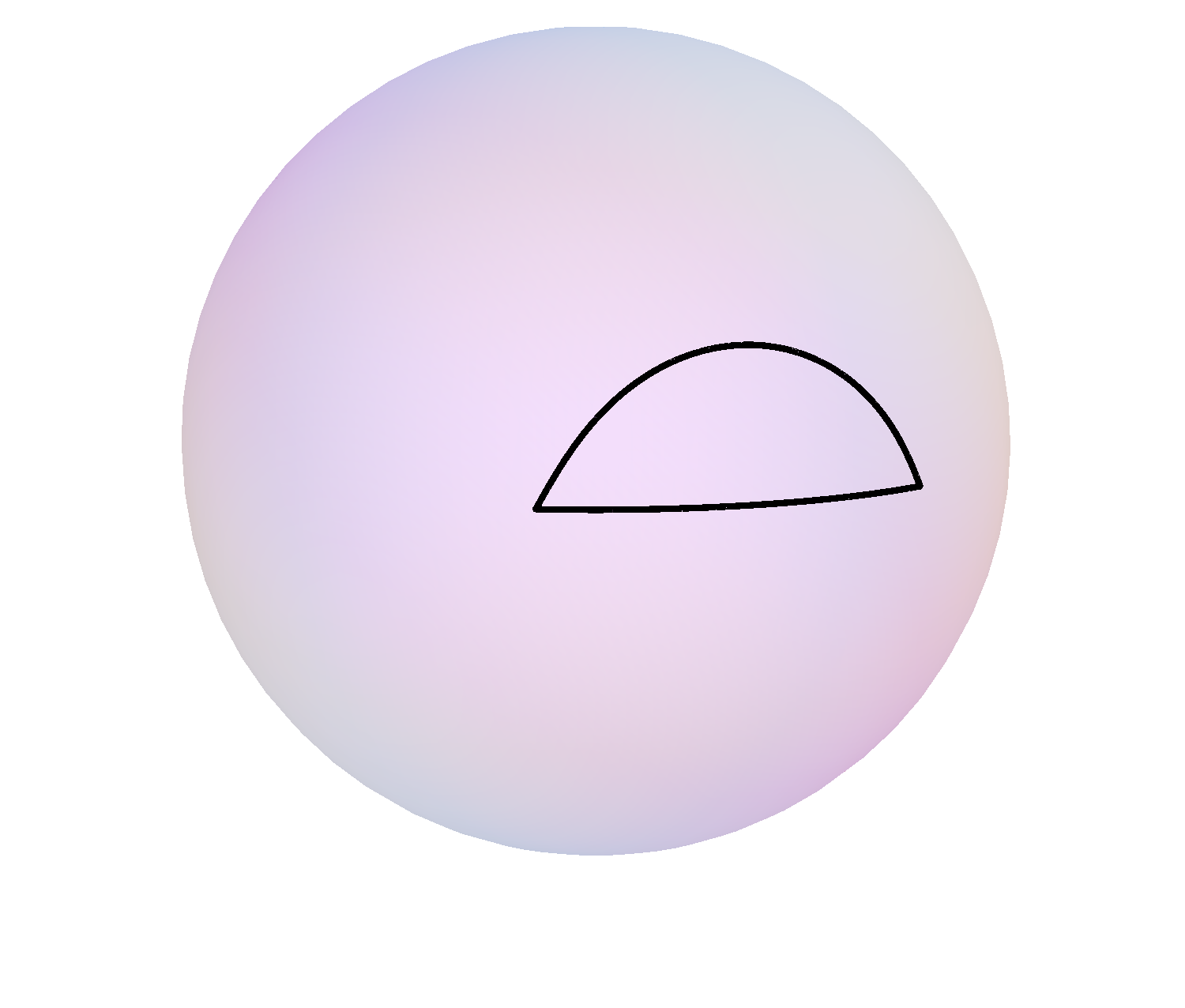}\\
\end{center}
\caption{Boundary curves for the petal solution. On the
  top row $\vec x|_{z=0}$ is plotted with $C=0.87$, $1.7$, $0.37$ and $b =
  0.57$, $3.41$, $1.37$ from left to right respectively. On the bottom
  row the curves on $S^5$, i.e. $\vec\theta|_{z=0}$ are given.}
\label{fig:petalex}
\end{figure}

The corresponding boundary $\vec\theta|_{z=0} = z\vec y|_{z=0}$ consists of a
curve ending at two points (these points are dual to the longitudes
of $\vec x|_{z=0}$) and then connected by a longitude (a piece of the
equator in this case), which is dual to the point of $\vec x|_{z=0}$
at the north pole, i.e. at $\s = \infty$. The petal solution\footnote{We show the
  $S^2\subset S^5$ upside down in figure \ref{fig:petal} to display the features of the solution
optimally.} is shown
in figure \ref{fig:petal} for $b=2$ and $C=\sqrt{21}/2$. The parameter
$C$ controls the extent to which the ``petal'' described by $\vec
x|_{z=0}$ descends away from the north pole. Specifically, the
longitudes extend from $x_3 = 1$ (i.e. the north pole) down to $x_3 =
1-(b^2-1)^2/(2 b^2 C^2)$. The parameter $b$ controls the opening
angle of the longitudes, given by $\pi (b^2-1)/(b^2+1)$. Several
examples of boundary curves are given in figure \ref{fig:petalex}.
%
\subsection{Area enclosed by  $\vec x|_{z=0}$}

It is a simple matter to evaluate the area on $S^2$ contained by the boundary
curve $\vec x|_{z=0}$, since two of the boundaries are
longitudes. Using standard spherical polar coordinates
\be
\vec x|_{z=0} = (\sin\vt \cos\vp, \sin\vt\sin\vp, \cos\vt),
\ee
the area is given by
\be
A_1 = \int_0^{\pi\frac{b^2-1}{b^2+1}} d\t \,\dot \vp(\t) \left(1- \cos\vt(\t)\right),
\ee
where $(\vt(\t),\vp(\t))$ describes the curve connecting the two
longitudes. One finds
\bsp
A_1 = - &\int_0^{\pi\frac{b^2-1}{b^2+1}} d\t \,
\frac{4(b^4-1)^2\sin^2\,\frac{b^2+1}{b^2-1}\t}{\left(1+b^4+2b^2
  \cos\,2\frac{b^2+1}{b^2-1}\t\right)}\\
&\times\frac{1}{
\Bigl[(b^2-1)^2 + 2(1+b^4) C^2 - (1+b^4-2b^2(1+2 C^2))\cos\,2\frac{b^2+1}{b^2-1}\t \Bigr]},
\end{split}
\ee
which is remarkably free of dependence on $b$, and evaluates to
\be
A_1 = 2\pi\left(1 - \frac{C}{\sqrt{1+C^2}}\right).
\ee
We may then verify (\ref{Cexp}), i.e.
\be
\frac{4\pi -2 A_1}{2\sqrt{(4\pi -A_1)A_1}} = C.
\ee
%

\subsection{Regularized area of the worldsheet}

According to (\ref{Sreg}) and (\ref{Cexp}) we expect that
\be\label{check}
S_{\text{reg.}} = \frac{\sqrt{\l}}{4\pi} \int d^2\s \left(\p_a \vec
\theta \cdot \p^a \vec \theta + \frac{1}{z} \nabla^2 z \right) =-\frac{ \sqrt{\lambda}}{\sqrt{1+C^2}}.
\ee
Owing to the complexity of $\vec \theta(\s,\t)$, the integral is very
hard to evaluate in a closed form. However numerical integration works
very well if the $b$ parameter is chosen appropriately. We have
verified (\ref{check}) over a wide selection of parameters using
numerical integration and have verified it with $10^{-6}$ percent-error accuracy.
\begin{figure}
\begin{center}
\includegraphics[height=1.65in]{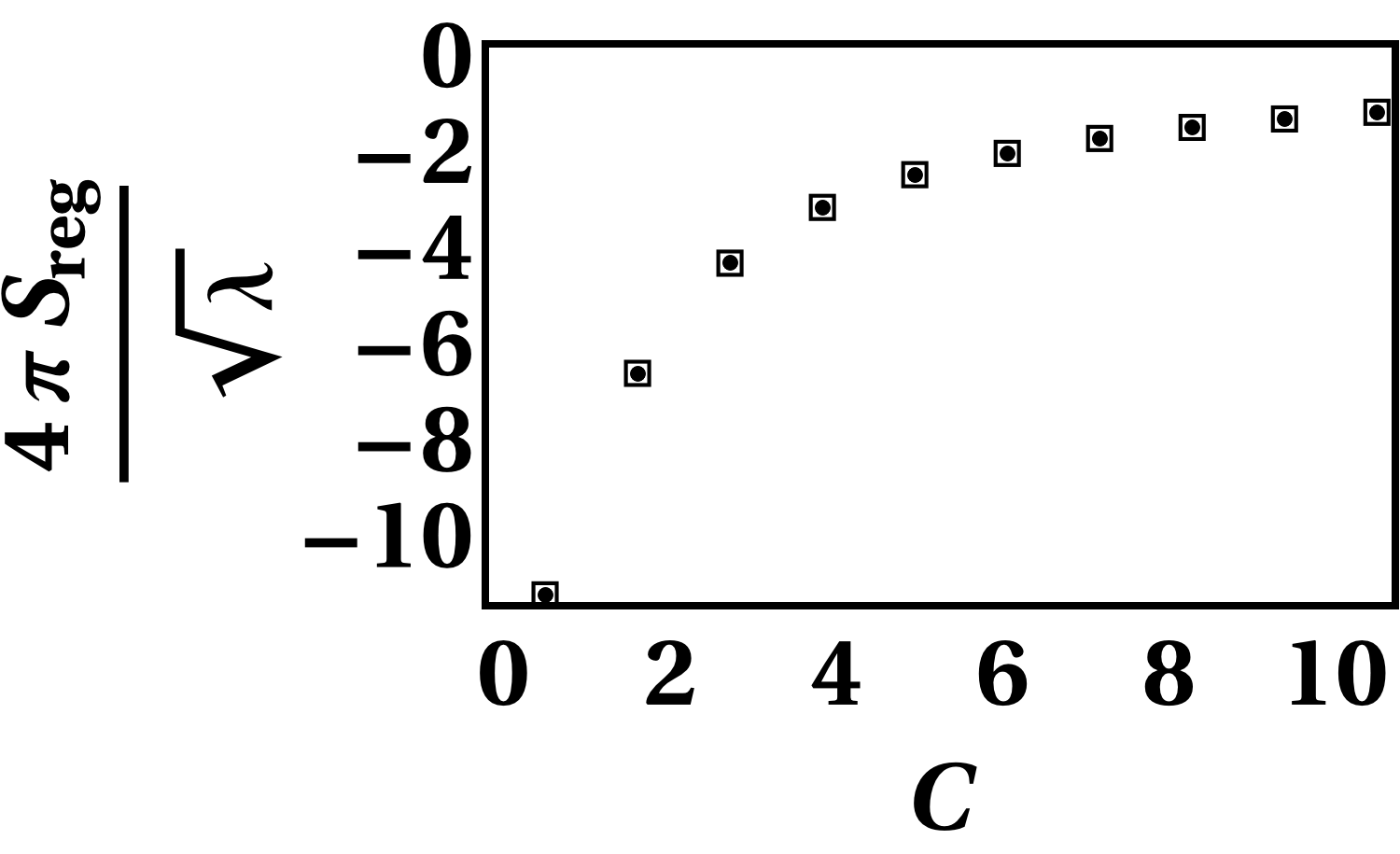} 
\includegraphics[height=1.65in]{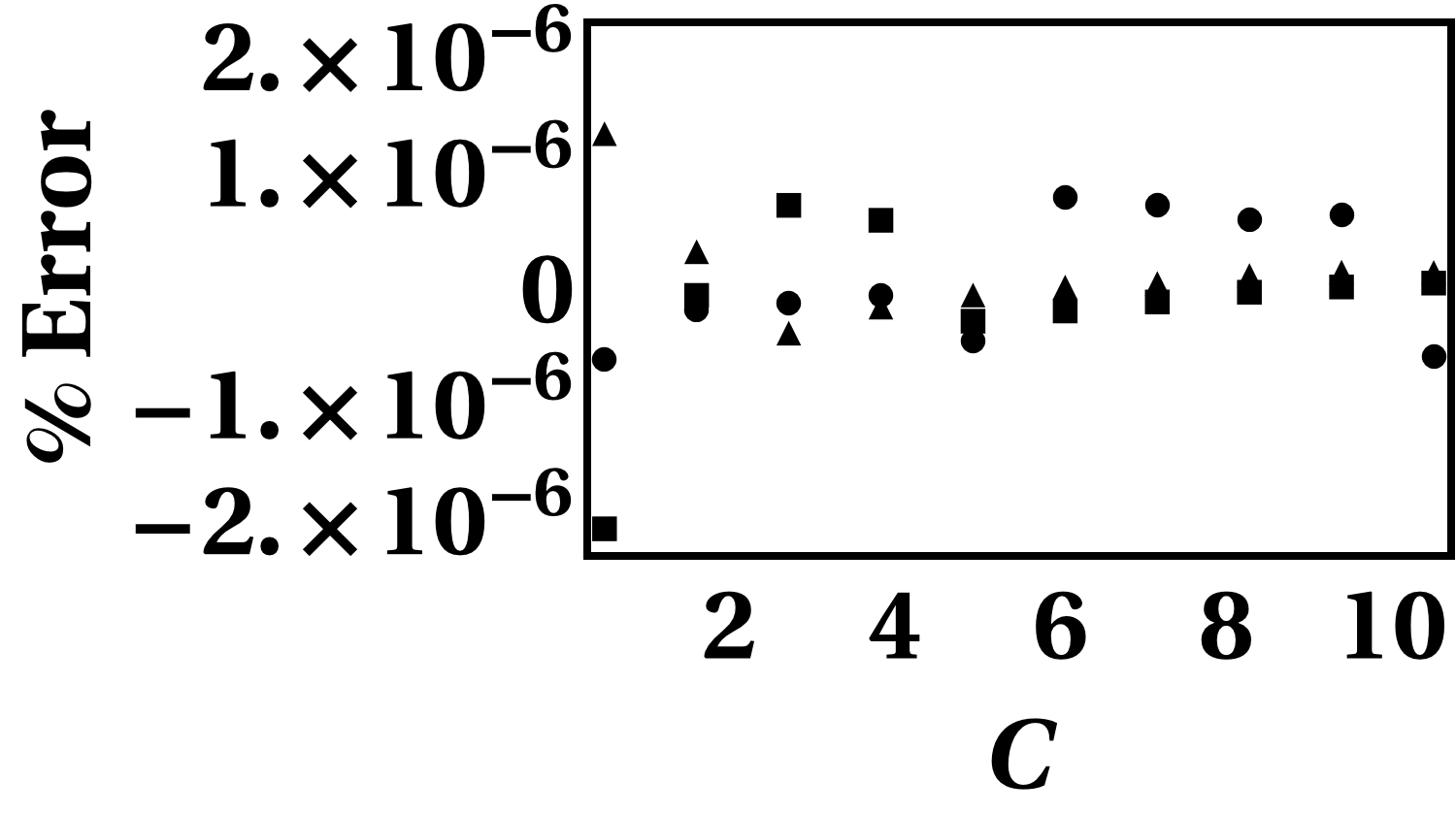} 
\end{center}
\caption{Results from numerical integration of (\ref{check}). On the
  left boxes are the prediction given on the RHS of (\ref{check})
  while the dots are the results of numerical integration of the
  LHS. Each data point represents three numerical integrations where
  $b$ is set to $\sqrt{1+C^2}$ and $\sqrt{1+C^2}\pm 3C/5$. The data
  are not distinguishable by their $b$ values in the plot, i.e. they
  lie atop one another. The percent error, defined in the text below,
  is given in the plot on the right: squares indicate
  $b=\sqrt{1+C^2}$, circles $b=\sqrt{1+C^2}-3C/5$, and triangles
  $b=\sqrt{1+C^2}+3C/5$.}
\label{fig:petalregarea}
\end{figure}
In figure \ref{fig:petalregarea} we show the results of numerical
integration performed using the Cuba package \cite{Hahn:2004fe} for a
range of values of $C$, and for each range three representative values
of $b$: $\sqrt{1+C^2}$ and $\sqrt{1+C^2}\pm3 C/5$. In the plot on the left the boxes are the
predicted value from the RHS of (\ref{check}) while the dots are the result of
numerical integration of the LHS (the data for each choice of $b$ lie
atop one another). The percent error defined as the difference
between prediction and numerical integration, divided by prediction,
and multiplied by $100$, are also plotted.

\section{Discussion}
\label{sec:disc}

Despite the mapping of the problem of finding string worldsheets
corresponding to 1/8-BPS Wilson loops on $S^2$ in ${\cal N}=4$ SYM to
an auxiliary $\s$-model on $S^3$, finding explicit solutions is
non-trivial. The reason is that the additional constraints
(\ref{addcons}) greatly constrain the solutions to the $\s$-model
which correspond to Wilson loops. In this paper we have found a
two-parameter family of solutions obtained by using the dressing
method on the known longitude solution.

It would be interesting to try the dressing method on different
starting solutions, such as the latitude solution, and also to try
dressing multiple times. Of course, it would also be very beneficial to
attempt to impose the additional constraints in a systematic way, but
this seems very difficult.

\section*{Acknowledgements}

CK has been supported by Deutsche Forschungsgemeinschaft via SFB
647. DY was supported by FNU through grant number 272-08-0329. We
would like to thank H.  Dorn, G. Jorjadze and J. Plefka for useful
discussions.

\bibliography{dresswl}%
\end{document}